\documentclass[aps,preprint,groupedaddress]{revtex4}
\usepackage{amssymb}
\usepackage{amsbsy}
\usepackage{rotating}
\usepackage{enumerate}
\usepackage{amsmath}
\usepackage{rotating}
\usepackage[latin1]{inputenc}
\usepackage{amsthm,mathrsfs,amsopn}
\DeclareMathOperator{\Tr}{Tr}

\usepackage{graphicx}

\begin{document}

\title{Stochastic Turing Patterns for systems with one diffusing species}

\author{Laura Cantini$^{1}$, Claudia Cianci$^{2}$, Duccio Fanelli$^{3}$, Emma Massi$^{1}$, Luigi Barletti$^{1}$}
\affiliation{
1. Dipartimento di Matematica e Informatica ``U.\ Dini'', University of Florence, Viale Morgagni 67/A, 50139 Florence, Italy\\
2. Dipartimento di Sistemi e Informatica and INFN, University of Florence, Via S. Marta 3, 50139 Florence, Italy\\
3. Dipartimento di Fisica e Astronomia, University of Florence and INFN, Via Sansone 1, 50019 Sesto Fiorentino, Florence, Italy}

\begin{abstract}
The problem of pattern formation in a generic two species reaction--diffusion model is studied, 
under the hypothesis that only one species can diffuse. For such a system, the classical Turing instability cannot take place. At variance, by working in the generalized setting of a stochastic formulation to the inspected problem, Turing like patterns can develop, seeded by finite size corrections.  General conditions are given for the stochastic Turing patterns to occur. The predictions of the  theory are tested for a specific case study.  
\end{abstract}

\maketitle

\vspace{0.8cm}

\section{Introduction} \label{S1}

Spatio temporal self-organized patterns \cite{Murray} can spontaneously emerge in a reaction-diffusion system. 
A small perturbation of a homogeneous fixed point can for example amplify, as follows a symmetry breaking instability seeded by diffusion, and eventually yield to a steady state non homogeneous 
solution. 
These are the Turing patterns \cite{Turing}, recurrently investigated in chemistry \cite{Belousov, Strogatz} and biology \cite{Murray}. 

The majority of studies devoted to the Turing instability consider two, mutually interacting, species. More specifically, and following the customarily accepted paradigm,
one species activates the production of the other, this latter acting through an inhibitor feedback. 
Systems of three \cite{satnoianu} simultaneously diffusing species have been also considered and shown to display a rich zoology of possible patterns and instabilities. 
Patterns can also develop if only one species is allowed to diffuse in the embedding medium, provided the system is composed of at least three coupled species  \cite{ermentrout}.  
In contrast, it is well known \cite{ermentrout} that two species systems where only one species can migrate, cannot undergo Turing instability. Models however exist which fall within this category \cite{goldbeter}. For this reason, it is of 
general interest to theoretically explore the possibility of bifurcation patterns of such systems, beyond the classical Turing framework. 
This paper aims at elaborating along these lines, by considering the generalized
concept of stochastically driven patterns.

Reaction-diffusion systems are in fact generally studied by resorting to deterministic mathematical models.
 The continuum concentrations of the interacting species is hence monitored over space and in time. 
As opposed to this, one can develop an individual based description of the scrutinized dynamics, which effectively accounts for
the inherent discreteness of the system. Stochastic contributions, stemming from finite size corrections, can thus modify the idealized mean field picture and occasionally 
return alternative scenarios to interpret available data.  

In a series of recent publications, the effect of the intrinsic noise was indeed shown to create stochastic patterns, in a region of the parameters for which macroscopically ordered structures do not occur.  
When the deterministic dynamics predicts a stable homogeneous state, the stochastic component can amplify via a resonant mechanism, giving birth to stochastic Turing patterns \cite{Goldenfeld, Biancalani_2010,Fanelli_2011,MainiWoolley}.  
The effect of finite size fluctuations can be characterized with numerical simulations, but also analytically with a mathematical technique, known as van Kampen system size expansion. 
This allows to expand the governing master equation, which accounts for the role of demographic fluctuations. 
At the first order of the expansion, the deterministic mean-field  model is obtained, while the second order contributions form an equation for the stochastic fluctuations. 

Working in this context, we will consider a simple birth and death model, with two species, of which one can diffuse. 
The reaction rates are assumed to be generic non linear functions of the concentration amount. Conditions for the emergence of stochastic Turing patterns are derived. More concretely, stochastic Turing patterns can materialize 
if the power spectrum of fluctuations has at least a peak for a non zero spatial wave number $k$ for $\omega$, the Fourier time frequency, equal to zero. 
We will here prove that a non trivial maximum of the power spectrum exists, if the system matches  
specific conditions that we shall mathematically characterize. The validity of our conclusions are tested for a simple non linear model, which falls in the general class of 
models inspected. With reference to this specific case study, we perform stochastic simulations through the Gillespie's algorithm and confirm a posteriori the adequacy of the  predictions.

The paper is organized as follows.
In the next section we will prove that, over a continuum support, the Turing instability cannot take place for reaction-diffusion models
with two interacting  species of which only one is allowed to diffuse  \cite{ermentrout}. 
If space is instead discrete, Turing like pattern can in principle take place, but only if the non diffusing species acts as a self-activator. 
However, when the condition for the instability are met, the most unstable mode $k$ is always located in $\pi$, a trivial consequence of the imposed discretization. 
As we shall here demonstrate, accounting for the intrinsic finite size fluctuations allows one to obtain a more complex landscape of possible instabilities. 
In Section \ref{S3} we introduce the stochastic birth and death model that we shall use as a reference case study. 
The model is completely  general and the reaction rates are assumed to depend on the species concentration, via generic non linear functions.  
Then, in Section \ref{S4}, we first derive the mean-field deterministic limit: the only request that we shall put forward has to do with the existence of a stable fixed point for the aspatial mean-field system. 
We then proceed to derive the Fokker-Planck equation that describes the fluctuations. 
From this, in Section \ref{S5}, we calculate the power spectrum of fluctuations, and find the mathematical conditions for having stochastic Turing patterns. 
We turn in Section \ref{S6} to considering a particular non-linear model, to verify the correctness of our predictions. 
Finally, in Section \ref{S7} we sum up and conclude. 

\section{Deterministic reaction-diffusion system with one diffusing species}  \label{S2}

Let us start by considering two species respectively characterized by the continuum concentrations $\phi(\mathbf{r},t)$ and $\psi(\mathbf{r},t)$. Here $\mathbf{r}$ stands for the spatial variable and $t$ represents time. Imagine the following general system to rule the dynamics of the concentrations:
\begin{eqnarray}
\frac{\partial \phi}{\partial t} &=& f(\phi, \psi) + D \nabla^2 \phi \nonumber \\ 
\frac{\partial \psi}{\partial t} &=& g(\phi, \psi)  
\label{syst_general}
\end{eqnarray}  
where $\nabla^2$ is the standard Laplacian operator and the functions $f(\cdot, \cdot)$ and $g(\cdot, \cdot)$ account for the interactions among the species. 
As anticipated we are focusing on the specific case study where just one species, specifically $\phi$, is allowed to diffuse, $D$ denoting its diffusion coefficient. 
Notice that $\psi$ is also function of the spatial variable $\mathbf{r}$, as it depends on the concentration $\phi$, the species which can in turn migrate. We shall here assume that a fixed point of the homogeneous system exists. This is a uniform solution $\phi(\mathbf{r})=\hat{\phi}$, $\psi(\mathbf{r})=\hat{\psi}$, with $\hat{\phi}$ and $\hat{\psi}$ constants, such that  $f(\hat{\phi}, \hat{\psi})= g(\hat{\phi}, \hat{\psi})=0$. We shall furthermore assume that the fixed point ($\hat{\phi}$, $\hat{\psi}$) is stable. In the following we will prove that no Turing instability can occur, if just one species can diffuse.  
\par
To this end we consider a small perturbation $\mathbf{w}$ of the initial homogeneous stationary state, in formulae:
\begin{equation}
\mathbf{w}=\begin{pmatrix}
\phi-\hat{\phi}\\
\psi-\hat{\psi}\\
\end{pmatrix}.
\end{equation}
Since $|\mathbf{w}|$ is by hypothesis small we can linearize system (\ref{syst_general}) around the fixed point
and so eventually obtain:
\begin{equation}
\dot{\mathbf{w}}=\mathcal{J}\mathbf{w}+\mathbf{D} \nabla^{2}\mathbf{w}, \hspace{1 cm} \mathbf{D}
= \begin{pmatrix} D  & 0\\ 0 & 0\\ \end{pmatrix}.
\label{eqxw}
\end{equation} 
where $\dot{\mathbf{w}}$ represents the time derivative of $\mathbf{w}$ and  $\mathcal{J}$ is the Jacobian matrix defined as: 
\begin{align}
 \mathcal{J}=\begin{pmatrix}
f_{\phi} & f_{\psi}\\
g_{\phi} & g_{\psi}\\
\end{pmatrix}, 
\label{jacobian}
\end{align}
where e.g.\ $f_{\phi}$ stands for  $ \partial f / \partial \phi$ evaluated at the fixed point $(\hat{\phi},\hat{\psi})$. 
Similar definitions apply to the other entries of the matrix $\mathcal{J}$. 
\par
To solve the above system (\ref{eqxw}), subject to specific boundary conditions,  one can introduce the eigenfunctions 
$\mathbf{W}_k(\mathbf{x})$ of the Laplacian, such that
$$
 - \nabla^2 \mathbf{W}_k(\mathbf{x}) = k^2 \mathbf{W}_k(\mathbf{x}), 
$$
for all $k \in \sigma$, where $\sigma$ is a suitable (unbounded) spectral set.
Then we expand
\begin{equation}   
\mathbf{w}(\mathbf{x},t)=\sum_{k \in \sigma}c_{k}e^{\lambda (k) t}\mathbf{W}_{k}(\mathbf{x}),
\label{soluz}
\end{equation}
where the constants $c_k$ refer to the initial condition. 
This is equivalent to performing a Fourier-like transform of the original equation. 
The function $\lambda (k)$, also called dispersion relation, controls the growth (or damping) of the perturbation. 
More specifically the solution of the linearized system (\ref{eqxw}) exists if
\begin{equation}
\det{(\lambda I-\tilde{\mathcal{J}})}=0 
\label{polcar}
\end{equation}
where $\det(\cdot)$ is the determinant and
\begin{equation} 
\tilde{\mathcal{J}}=
\begin{pmatrix}
f_{\phi} -Dk^2      &   f_{\psi} \\
g_{\phi}      &  g_{\psi}
\end{pmatrix}.
\label{jac}
\end{equation}
A simple calculation yields:
\begin{equation}
\lambda(k)=\frac{(\Tr{\mathcal{J}}-Dk^2)+\sqrt{(\Tr{\mathcal{J}}-Dk^2)^2-4(\det{\mathcal{J}}-Dk^{2}g_{\psi})}}{2}
\label{dT}
\end{equation}
where $\Tr(\cdot)$ denotes the trace. Since we are interested in the growth of unstable perturbations, we have here selected the largest $\lambda(k)$.  The Turing instability occurs if one can isolate a finite domain in $k$ for which $\lambda(k)>0$. In formulae: 

\begin{equation*}
(\Tr{\mathcal{J}}-Dk^2)+\sqrt{(\Tr{\mathcal{J}}-Dk^2)^2-4(\det{\mathcal{J}}-Dk^{2}g_{\psi})}>0
\end{equation*} 
\begin{equation*}
\implies 
\sqrt{(\Tr{\mathcal{J}}-Dk^2)^2-4(\det{\mathcal{J}}-Dk^{2}g_{\psi})}> -(\Tr{\mathcal{J}}-Dk^2)
\end{equation*}
\begin{equation*}
\implies -4(\det{\mathcal{J}}-Dk^{2}g_{\psi})>0 
\end{equation*}
\begin{equation}
\implies Dk^{2}g_{\psi}>\det{\mathcal{J}}.
\label{last}
\end{equation} 

The right hand side contribution in equation \eqref{last} is positive as the homogeneous fixed point is supposed to 
be stable. If  $g_{\psi}<0$ it is clear that  \eqref{last} does not admit solutions, the left hand side of the equation being negative.  At variance, when $g_{\psi}>0$ we have:
\begin{equation}
k^{2}>\frac{\det{\mathcal{J}}}{Dg_{\psi}}\; \implies k<-\sqrt{\frac{\det{\mathcal{J}}}{Dg_{\psi}}}\quad \text{and}\quad k
>\sqrt{\frac{\det{\mathcal{J}}}{Dg_{\psi}}}.
\label{gvpos}
\end{equation}
Equation \eqref{gvpos} implies that the relation of dispersion $\lambda(k)$ is positive for all values of $k\in \sigma$
above a critical threshold $k_c=\sqrt{\det{\mathcal{J}}/(Dg_{\psi})}$. 
The quantity $\lambda(k)$ grows as $k$ does, the instability involving smaller and smaller spatial scales. 
It is therefore not possible to delimit a finite window in $k$ for which $\lambda(k)$ is found to be positive, and, hence,
the Turing instability cannot take place. In conclusion, we have here confirmed a well establish fact \cite{ermentrout}: 
a two species systems where only one species can migrate, cannot undergo Turing instability. 

\par

Let us now turn to considering the case where the spatial support is supposed to be discrete. 
In practice, this amounts to assume the physical space, in any dimension, to be partitioned in 
a large collection of mesoscopic patches, where the constituents are assumed to be uniformly mixed. 
The diffusion can take place between adjacent patches. 
The differential equations that govern the evolution of the concentration are therefore discrete in space, a setting 
that is for instance of interest when  reaction-diffusion models are applied to ecology \cite{McKaneEcologyTrends}.  

For simplicity, and without losing generality, we will hereafter consider the problem in one dimension,  
assuming the physical space to be segmented in $\Omega$ cells, each of finite linear size $a$. 
We label $\phi_i$ and $\psi_i$, with $i = 1,\ldots,\Omega$, the discrete concentrations, that respectively replace 
their continuum analogues  $\phi$ and $\psi$. 

The discrete Laplacian operator $\Delta$ is defined as:
\begin{equation} 
\Delta \phi_i = \frac{1}{a^2} \sum_{j=i \pm 1} \left( \phi_j - \phi_i \right)
\end{equation}
and periodic boundary conditions  at $i = 1$ and $i = \Omega$ will be assumed throughout the rest of the paper.
Let $\delta$ denote the transition probability per unit of time that control the migration between neighbors 
mesoscopic patches.
In the continuum limit $\delta a^2 \rightarrow D$,  when $a \rightarrow 0$.
The discrete reaction diffusion system can be therefore written as: 
\begin{equation}
\left\{
\begin{alignedat}{1}
&\frac{\partial \phi_{i}}{\partial t}=f\left( \phi_{i}, \psi_{i}\right)+\left (\delta a^2 \right)  \Delta \phi_i \\
&\frac{\partial \psi_{i}}{\partial t}=g(\phi_{i},\psi_{i})\label{R-D1d}.
\end{alignedat}
\right.
\end{equation}

To study the onset of the instability, we operate in analogy with what has been done above and 
perform a spatio--temporal Fourier transform of eqs. (\ref{R-D1d}). 
The transform of the discrete Laplacian $\Delta$ reads $\tilde{\Delta}_k= (2/a^2)(\cos(ak)-1)$.  
Proceeding in the analysis, one ends up with the following relation of dispersion: 
\begin{equation} 
\lambda(k)=\frac{h(k)+\sqrt{h(k)^2-4(\det{\mathcal{J}}+2 \delta (\cos(ak)-1))g_{\psi}}}{2}
\label{dTd}
\end{equation}
where $h(k)=\Tr{\mathcal{J}}+2 \delta (\cos(ak)-1)$. 
By imposing $\lambda(k)>0$ one obtains, after a simple algebraic manipulation, the following condition: 
\begin{equation} 
\delta (1-\cos(ak))g_{{\psi}}>2\det{\mathcal{J}}.
\label{disc}
\end{equation} 
As it happens for the  case of the continuum,  no solution of  \eqref{disc} are possible when $g_{{\psi}}<0$, 
namely when the non diffusing species has a self-inhibitory effect.
At variance, if $g_{{\psi}}>0$  a finite interval in $k$ can be found where $\lambda(k)$ is different from zero, and the system can therefore experience a Turing instability which is indeed seeded by the discreteness of the spatial support. The most unstable mode $k_M$ is however found to be $k_M=\pi/(2 a)$, a trivial solution which stems from having assumed a discrete spatial support. It is worth emphasizing that, as expected, $k_M$ diverges to infinity when the size of the patch $a$ goes to zero \footnote{The fact a discretised domain can produce wave modes to appears that do not exist in the continuum case was also noticed in \cite{MainiWoolley}.}. 
\par
Starting from this setting, we will work in the context of a stochastic formulation of the generic reaction diffusion system considered above and show that finite size corrections 
can eventually drive the emergence of Turing like patterns. We will in particular specialize on the case of a model defined on a discrete lattice and assume $g_{{\psi}}<0$. Under this  condition the Turing patterns cannot develop in the mean-field approximation.

\section{The Model and its Master Equation}  \label{S3}

The system that we are going to study is a general two species birth-death model, in which one of the species diffuses.
As already mentioned, we assume the physical space to be partitioned in $\Omega$ patches \footnote{For the sake of simplicity, and without loosing generality we will set $a=1$ in the following.}, 
and label with $V$ their carrying capacity. 
The integer index $i$ runs from $1$ to $\Omega$ and identifies the cell to which the species belong.
Label the two species $Z$ and $Y$ and assume the following chemical reaction scheme:
\begin{equation}
\begin{split}
&Z_{i}  \xrightarrow[]{\;\;\alpha\;\;} Z_{i}+1 \hspace{2 cm} \alpha= \frac{1}{\Omega}\frac{V}{s_i}f_{1}\bigg(\frac{s_i}{V},\frac{q_i}{V}\bigg)\\
&Z_{i}  \xrightarrow[]{\;\;\beta\;\;} Z_{i}-1\hspace{2 cm} \beta=\frac{1}{\Omega}\frac{V}{s_i}f_{2}\bigg(\frac{s_i}{V},\frac{q_i}{V}\bigg)\\
&Y_{i} \xrightarrow[]{\;\;\gamma\;\;} Y_{i}+1\hspace{2.15 cm} \gamma=\frac{1}{\Omega}\frac{V}{q_i}g_{1}\bigg(\frac{s_i}{V},\frac{q_i}{V}\bigg)\\
&Y_i \xrightarrow[]{\;\;\rho\;\;} Y_{i}-1\hspace{2.15 cm} \rho=\frac{1}{\Omega}\frac{V}{q_i}g_{2}\bigg(\frac{s_i}{V},\frac{q_i}{V}\bigg)
\end{split}
\label{EC1}
\end{equation}  
We indicated as $s_i$ the number of elements of species $Z$ and with $q_i$ the number of elements of species 
$Y$ in the cell $i$. 
Moreover, we require that $f_{1}, f_{2}, g_{1}, g_{2}$ are sufficiently regular functions of the discrete number
concentrations $s_i/V$ and $q_i/V$.  
\par
We assume that only $Z$ diffuses and therefore write
\begin{equation}
Z_{i}\xrightarrow[]{\;\delta/w\Omega\;} Z_{j} \hspace{1.8 cm} Z_{j}
\xrightarrow[]{\;\delta/w\Omega\;} Z_{i}, \hspace{1 cm} \text{with}\;  j \in \{i-1,i+1\},
\label{EC2}
\end{equation}
where, in general, $w$ is the number of neighboring cells of a given cell $i$ and, therefore, $w=2$ in the present
one-dimensional case. 
A state of the system is characterized by two vectors, respectively $\vec{s}=(s_1, s_2, ..., s_\Omega)$ and 
$\vec{q}=(q_1, q_2, ..., q_\Omega)$. 
It is worth emphasizing that the model is completely general: virtually any system composed by two species, 
one of each diffusing, can be cast in the form introduced above, upon a proper choice of the functions 
$f_{1}, f_{2}, g_{1}, g_{2}$.
\par
We then turn to write down the master equation that governs the dynamics of the system. To this end we need to calculate the transition probability associated with each reaction:
\begin{align*}
&T(s_{i}+1,q_i|s_i,q_i)=\alpha \frac{s_i}{V}\\
&T(s_{i}-1,q_i|s_i,q_i)=\beta \frac{s_i}{V}\\
&T(s_{i},q_i+1|s_i,q_i)=\gamma \frac{q_i}{V}\\
&T(s_{i},q_i-1|s_i,q_i)=\rho \frac{q_i}{V}\\
&T(s_{i}+1,s_{j}-1|s_i,s_j)=\frac{\delta}{\Omega}\frac{s_j}{bV}\\
&T(s_{i}-1,s_{j}+1|s_i,s_j)=\frac{\delta}{\Omega}\frac{s_i}{bV}.
\end{align*}
By introducing the following ``step operators'':
$$
  \varepsilon^{\pm}_{s_i}f(\vec{s},\vec{q})=f(\ldots,s_i \pm 1,\ldots,\vec{q}),
  \qquad  
  \varepsilon^{\pm}_{q_i}f(\vec{s},\vec{q})=f(\vec{s},\ldots,q_i \pm 1,\ldots),
$$
the master equation reads:
\begin{equation}
\begin{aligned}
\frac{d}{dt}P(\vec{s},\vec{q},t) &= \sum_{i=1}^{\Omega}\bigg[\left( \varepsilon^+_{s_i}-1\right)T(s_i -1,q_i|s_i,q_i)
+ \left(\varepsilon^-_{s_i}-1\right)T(s_i +1,q_i|s_i,q_i)
\\
&+\left( \varepsilon^+_{q_i}-1\right)T(s_i,q_i-1|s_i,q_i) + \left(\varepsilon^-_{q_i}-1\right)T(s_i,q_i+1|s_i,q_i)\bigg]
P(\vec{s},\vec{q},t)
\\
&+\sum_{i=1}^{\Omega}\sum_{j\in\{i-1,i+1\}}\bigg[\left(\varepsilon_{s_i}^{+}\varepsilon_{s_j}^{-}-1\right)T(s_i -1,s_j+1|s_i,s_j)
\\
&+\left( \varepsilon^{-}_{s_i} \varepsilon^{+}_{s_j}-1\right)T(s_i +1,s_j-1|s_i,s_j)\bigg] P(\vec{s},\vec{q},t)
\end{aligned}
\label{MEfinale}
\end{equation}
where, in accordance with our assumption of periodic boundary conditions, we adopt a periodic convention for the indices out of the set $\{1,\ldots \Omega\}$.
\par
The master equation is difficult to handle analytically and we perform a van Kampen system size expansion, a perturbative calculation that introduces, by an ansatz, the following change of variables in the master equation: 
\begin{equation}\label{ip}
\frac{s_{i}}{V}=\phi_i+\frac{\xi_{i}}{\sqrt{V}}, \qquad
\frac{q_{i}}{V}=\psi_i+\frac{\eta_{i}}{\sqrt{V}}.
\end{equation}
The number density $s_i/V$  splits into two independent contributions: $\phi_i$ stands for the deterministic (mean-field)
concentration as measured in correspondence of the site $i$, and $\xi_i$ is a stochastic variable that quantifies  the 
fluctuation that perturbs the mean-field solution $\phi_i$. Similar considerations apply to $q_i/V$. 
The factor $1/\sqrt{V}$ takes into account the finite volume of the system. 
In the limit for infinite systems size, the fluctuations can be neglected and the stochastic system as formulated above converges
 to its deterministic analogue.  
 When working at finite $V$, stochastic fluctuations are important. 
The role of fluctuations can be quantitatively studied by implementing the aforementioned perturbative analysis, the van Kampen
expansion \cite{Vankampen}, which assumes the amplitude factor $1/\sqrt{V}$ to act as a small parameter. 
To this end we introduce the van Kampen hypothesis into the master equation and split the contributions of order 
$1/\sqrt{V}$ and $1/V$, to respectively obtain the mean field equation and Fokker-Planck equation. 
To carry out the calculation explicitly one needs to expand the functions $f_{1}, f_{2}, g_{1}, g_{2}$ with respect to the small 
parameter $1/\sqrt{V}$. 
As a representative example, we consider $f_1$ and obtain:  
\begin{equation}
f_1\left(\phi_i+\frac{\xi_i}{\sqrt{V}},\psi_i+\frac{\eta_i}{\sqrt{V}}\right)\approx f_{1}(\phi_i,\psi_i)+\frac{1}{\sqrt{V}}\frac{\partial f_{1}}{\partial \phi_i}(\phi_i,\psi_i)\xi_i+\frac{1}{\sqrt{V}}\frac{\partial f_{1}}{\partial \psi_i}(\phi_i,\psi_i)\eta_i + \cdots
\label{sviluppof}
\end{equation}
where the derivatives are evaluated at $\xi_{i}=0, \eta_{i}=0$. 
Similar results hold for $f_{2}$, $g_{1}$ and  $g_{2}$.
\par
Let us introduce the new distribution
\begin{equation}
\Pi(\xi_i,\eta_i,t)=P(s_i(\phi_{i}(t),\xi_i),q_i(\psi_{i}(t),\eta_i),t),
\end{equation}
where $s_i(\phi_{i}(t),\xi_i)$ and $q_i(\psi_{i}(t),\eta_i)$ are given by \eqref{ip}.
Inserting into the master equation, and expanding the step operators to second order, one eventually obtains
\begin{equation}
\sum_{i=1}^{\Omega}\frac{\partial \Pi}{\partial t}-\frac{\partial \Pi}{\partial \xi_i}\sqrt{V}\dot{\phi_i}-\frac{\partial \Pi}{\partial \eta_i}\sqrt{V}\dot{\psi_i}=[A+B+C]\Pi
\label{MEPI2fin}
\end{equation}
where the contributions $A, B, C$ take the following form:
\begin{align*}
A=\frac{1}{\Omega}\sum_{i=1}^{\Omega}&\Bigg\{\frac{1}{\sqrt{V}}\bigg[(f_{2}-f_1)\frac{\partial}{\partial \xi_i}\bigg]+\\
&+\frac{1}{V}\bigg[\frac{\partial}{\partial \xi_i}\bigg(\frac{\partial f_{2}}{\partial \phi_i}-\frac{\partial f_{1}}{\partial \phi_i}\bigg)\xi_i+\frac{\partial}{\partial \xi_i}\bigg(\frac{\partial f_{2}}{\partial \psi_i}-\frac{\partial f_{1}}{\partial \psi_i}\bigg)\eta_i+\frac{1}{2}\left(f_{1}+f_2\right)\frac{\partial^2}{\partial \xi_{i}^2}\bigg]\Bigg\},
\end{align*}
\begin{align*}
B=\frac{1}{\Omega}\sum_{i=1}^{\Omega}&\Bigg\{\frac{1}{\sqrt{V}}\bigg[(g_{2}-g_1)\frac{\partial}{\partial \eta_i}\bigg]+\\
&+\frac{1}{V}\bigg[\frac{\partial}{\partial \eta_i}\bigg(\frac{\partial g_{2}}{\partial \phi_i}-\frac{\partial g_{1}}{\partial \phi_i}\bigg)\xi_i+\frac{\partial}{\partial \eta_i}\bigg(\frac{\partial g_{2}}{\partial \psi_i}-\frac{\partial g_{1}}{\partial \psi_i}\bigg)\eta_i+\frac{1}{2}\left(g_{1}+g_2\right)\frac{\partial^2}{\partial \eta_{i}^2}\bigg]\Bigg\},
\end{align*}
\begin{align*}
C=&\frac{\delta}{b\Omega}\sum_{i=1}^{\Omega}\sum_{j\in\{i-1,i+1\}}\Bigg\{\frac{1}{\sqrt{V}}\bigg[\bigg(\frac{\partial}{\partial \xi_i}-\frac{\partial}{\partial \xi_j}\bigg)\phi_{i}+\bigg(\frac{\partial}{\partial \xi_j}-\frac{\partial}{\partial \xi_i}\bigg)\phi_{j}\bigg]\;+\hspace{5 cm}\\
&+\frac{1}{V}\Bigg[\bigg(\frac{\partial}{\partial \xi_i}-\frac{\partial}{\partial \xi_j}\bigg)\xi_{i} + \bigg(\frac{\partial}{\partial \xi_{j}}-\frac{\partial}{\partial \xi_i}\bigg)\xi_{j}+\frac{1}{2}\bigg(\frac{\partial^2}{\partial \xi_{i}^2}+\frac{\partial^2}{\partial \xi_{j}^2}-2\frac{\partial}{\partial \xi_i}\frac{\partial}{\partial \xi_j}\bigg)(\phi_{i}+\phi_{j}) \Bigg]\Bigg\}.
\end{align*}

\section{Equations for the mean-field and the fluctuations}  \label{S4}

Introducing the rescaled time variable $\tau \rightarrow t/\Omega V$, we obtain from \eqref{MEPI2fin} at the order $1/\sqrt{V}$
the following system of ordinary differential equations for the mean field concentrations $\phi_i$ and $\psi_i$:
\begin{equation}
\left\{
\begin{alignedat}{1}
&\dot{\phi}_i=f_{1}(\phi_i,\psi_i)-f_{2}(\phi_i,\psi_i)+\delta \triangle\phi_i\\
&\dot{\psi}_i=g_{1}(\phi_i,\psi_i)-g_{2}(\phi_i,\psi_i)
\end{alignedat} \label{R-D1d_from_Stoc}
\right.
\end{equation}
where $\triangle=(\phi_{i+1}-2\phi_{i}+\phi_{i-1})$, the discrete Laplacian for $a=1$.
To proceed in the analysis we suppose that the homogeneous system:
\begin{equation}
\left\{
\begin{alignedat}{1}
&\dot{\phi}_i=f_{1}(\phi_i,\psi_i)-f_{2}(\phi_i,\psi_i) \equiv f(\phi_i,\psi_i) \\
&\dot{\psi}_i=g_{1}(\phi_i,\psi_i)-g_{2}(\phi_i,\psi_i) \equiv g(\phi_i,\psi_i)
\end{alignedat}
\right.
\end{equation}
 admits a fixed stable point $(\hat{\phi},\hat{\psi})$. Notice that system (\ref{R-D1d_from_Stoc}), derived from a microscopic stochastic formulation,
 coincides with the general mean-field model (\ref{R-D1d}) considered in Section \ref{S2}.  
\par
\medskip





The Fokker Planck equation that describes the dynamics of the fluctuations is obtained by considering the terms proportional 
to $1/V$ in the master equation and reads as follows:
\begin{equation}\label{FPfinale}
\frac{\partial}{\partial \tau}\Pi=\sum_{i=1}^{\Omega}\left(-\sum_{r=1}^{2}\frac{\partial}{\partial\zeta_{r,i}}\left( \sum_{m=1}^2 \mathcal{J}_{rm,i} {\zeta_{m,i}}\Pi \right)+\frac{1}{2}\sum_{r,l=1}^{2}\sum_{j=i-1}^{i+1}\frac{\partial}{\partial\zeta_{l,i}}\frac{\partial}{\partial\zeta_{r,j}}\left(\mathcal{B}_{rl,j}^{(i)}\Pi\right)\right).
\end{equation}
Let us indicate as $\vec{\zeta_i}=({\zeta_{1,i}}, {\zeta_{2,i}})$ the vector $(\xi_{i},\eta_{i})$ in \eqref{FPfinale}. 
The $2 \times 2$ matrices $\mathcal{J}_i = \mathcal{J}_{rm,i}$ are given by
\begin{equation}
\label{Ji}
\mathcal{J}_i=
\begin{pmatrix}
\dfrac{\partial f_1}{\partial \phi_i}-\dfrac{\partial f_2}{\partial \phi_i}+\delta\triangle \quad&\quad \dfrac{\partial f_1}{\partial \psi_i}-\dfrac{\partial f_2}{\partial \psi_i}\\
&\\
\dfrac{\partial g_1}{\partial \phi_i}-\dfrac{\partial g_2}{\partial \phi_i}\quad &\quad \dfrac{\partial g_1}{\partial \psi_i}-\dfrac{\partial g_2}{\partial \psi_i},
\end{pmatrix}
\end{equation}
and the three-vectors $\mathcal{B}_{rl}^{(i)}$ are given by 
\begin{equation}
\begin{split}
\mathcal{B}_{11}^{(i)}&=\left(-\delta(\phi_i+\phi_{i-1})\mathbb{,}\delta(\phi_{i-1}+2\phi_i+\phi_{i+1})+f_{1}(\phi_i,\psi_i)+f_{2}(\phi_i,\psi_i)\mathbb{,}-\delta(\phi_i+\phi_{i+1})\right)\\
&\\
\mathcal{B}_{12}^{(i)}&=\mathcal{B}_{21}^{(i)}=(0,0,0), \hspace{1.6 cm} \mathcal{B}_{22}^{(i)}=(0,g_{1}(\phi_i,\psi_i)+g_{2}(\phi_i,\psi_i),0).
\end{split}
\label{BB}
\end{equation} 
Note that, in the above expressions, the indices $r$ and $l$ label the species while the indices $i$ and $j$ refer to the cells.
The matrix $\mathcal{J}_i$ is the Jacobian matrix of $(\phi_i,\psi_i) \mapsto (f_1-f_2,g_1-g_2)$, 
modified with the inclusion of the spatial contribution represented by the discrete Laplacian. 

Matrix $\mathcal{B}$ can be  cast in the more compact form:

\begin{equation}
\label{eqB}
\mathcal{B}_{rl,j}^{(i)}=\big(b^{(0)}_{rl}\delta_{i-j,0}+b^{(1)}_{rl}\delta_{|i-j|,1}\big)+b^{(1)}_{rl}\triangle
\end{equation}

where:

\begin{align*}
& b^{(0)}=
\begin{pmatrix}
2 \delta \hat{\phi}+ f_{1}(\phi_i,\psi_i) + f_{2}(\phi_i,\psi_i)   & 0\\
0 & g_{1}(\phi_i,\psi_i)+g_{2}(\phi_i,\psi_i)
\end{pmatrix}\\
&b^{(1)}=
\begin{pmatrix}
-\delta \phi_i \hspace{.5 cm} & 0\\
\hspace{.01 cm} 0 & 0
\end{pmatrix}
\end{align*}

We are interested in studying the fluctuations around the fixed point, when the deterministic system is in a steady state, i.e. when $(\phi_i,\psi_i) \equiv (\hat{\phi},\hat{\psi})$, $\forall i$. A powerful mean of investigation is the power spectrum of fluctuations, that allows us to resolve the typical spatio-temporal frequencies that are represented in the recorded signal. The analysis of the power spectrum is carried out in the next section.

\section{Power Spectrum of fluctuations}  \label{S5}
\label{sec_ps}

The above Fokker-Planck equation is equivalent \cite{Vankampen} to the Langevin equation: 
\begin{equation}
\frac{d}{dt}\zeta_{r,i}(t)=\sum_{l=1}^{2}\mathcal{J}_{rl,i}\zeta_{l,i}(t)+\lambda_{r,i}(t)
\label{Langevin}
\end{equation}
where $\lambda_{r,i}(t)$ is a stochastic contribution which satisfies the following relations:
\begin{align}
&\left<\lambda_{l,i}(t),\lambda_{r,i'}(t')\right>=\mathcal{B}_{lr,|i-i'|}\delta(t-t'), \label{autocorrelazione}
\\[6pt]
&\left<\lambda_{l,i}(t)\right>=0.\label{medianulla}
\end{align}
and $\left< \cdot \right>$ denotes expectation.
Upon Fourier transform one gets:
\begin{equation}
-i\omega\tilde{\zeta}_{r,k}(\omega)=\sum_{l=1}^{2}\tilde{\mathcal{J}}_{rl,k}\tilde{\zeta}_{l,k}(\omega)+\tilde{\lambda}_{r,k}(\omega)
\label{TFLangevin}
\end{equation}
where $\tilde{(\cdot)}$ stands for the Fourier transform both in space and time. 
Notice that matrix $\tilde{\mathcal{J}}_i$ coincides with the matrix $\mathcal{J}_i$ given in \eqref{Ji}
where the discrete Laplacian $\triangle$, is replaced by its Fourier transform $\tilde{\triangle}_k$. As previously remarked, and recalling that $a=1$, one gets: 
\begin{equation}
\tilde{\boldsymbol{\triangle}}_k=2(\cos(k)-1)
\label{TFlaplaciano}
\end{equation}
Define 
$$
  {\Phi}_{rl,k}(\omega)=-i\omega\delta_{rl}-\tilde{\mathcal{J}}_{rl,k},
$$ 
then the solution of \eqref{TFLangevin} reads:
\begin{equation}
\tilde{\zeta}_{r,k}(\omega)=\sum_{l=1}^{2}{\Phi}^{-1}_{rl,k}(\omega)\tilde{\lambda}_{r,k}(\omega).
\label{TF}
\end{equation}
The power spectrum of the stochastic variable $\zeta_{r,i}(t)$ is defined as:
\begin{equation}
P_{r}(k,\omega)=\left<|\tilde{\zeta}_{r,k}(\omega)|^{2}\right>.
\label{Ps}
\end{equation}

Making use of condition (\ref{autocorrelazione}) one gets:  

\begin{equation}
P_{r}(k,\omega)=\left<|\tilde{\zeta}_{r,k}(\omega)|^{2}\right>=\sum_{l,p=1}^{2}{\Phi}^{-1}_{rl,k}(\omega)\tilde{\mathcal{B}}_{lp,k}({\Phi}^{\dag})^{-1}_{rp,k}(\omega).
\label{PSfinale}
\end{equation}

By recalling expression (\ref{eqB}) one gets:

\begin{equation}
\tilde{\mathcal{B}}_{lp,k}=\big(b^{(0)}_{lp}+2b^{(1)}_{lp}\big)+b^{(1)}_{lp}\tilde{\triangle}_{k},
\label{TFautocorrelazione}
\end{equation}

which allows us to rewrite the power spectra in the form $P_{r}(k,\omega)$ \cite{McKane,De_Anna}:

\begin{equation}
P_{Z}(k,\omega)\equiv P_{1}(k,\omega)=\frac{C_{Z,k}+\tilde{\mathcal{B}}_{11,k}\omega^{2}}{(\omega^{2}-\Omega^{2}_{0})^2+\Gamma^{2}\omega^{2}},\label{Pz}
\end{equation}
\begin{equation}
P_{Y}(k,\omega)\equiv P_{2}(k,\omega)=\frac{C_{Y,k}+\tilde{\mathcal{B}}_{22,k}\omega^{2}}{(\omega^{2}-\Omega^{2}_{0})^2+\Gamma^{2}\omega^{2}}. \label{Py}
\end{equation}
where the functions $C_{Z,k}$ and $C_{Y,k}$ are respectively defined: 
\begin{equation}
\begin{split}
C_{Z,k}&=\tilde{\mathcal{B}}_{11,k}(\tilde{\hat{\mathcal{J}}}_{22,k})^{2}+\tilde{\mathcal{B}}_{22,k}(\tilde{\hat{\mathcal{J}}}_{12,k})^{2}-2\tilde{\mathcal{B}}_{12,k}\tilde{\hat{\mathcal{J}}}_{12,k}\tilde{\hat{\mathcal{J}}}_{22,k},\\
&\\
C_{Y,k}&=\tilde{\mathcal{B}}_{22,k}(\tilde{\hat{\mathcal{J}}}_{11,k})^{2}+\tilde{\mathcal{B}}_{11,k}(\tilde{\hat{\mathcal{J}}}_{21,k})^{2}-2\tilde{\mathcal{B}}_{12,k}\tilde{\hat{\mathcal{J}}}_{21,k}\tilde{\hat{\mathcal{J}}}_{11,k}
\end{split}
\end{equation}
and
\begin{eqnarray}
\Omega_{0} &=& \sqrt{\det\mathcal{\hat{J}}_{rl,k}}\\
\Gamma &=& -\Tr{\mathcal{\hat{J}}_{rl,k}}.
\end{eqnarray}

In the above expression, the symbol $\mathcal{\hat{(\cdot)}}$ indicates that from hereon the matrices are evaluated at the fixed point $(\hat{\phi},\hat{\psi})$; $\tilde{(\cdot)}$ stands instead for the spatial Fourier transform.

As anticipated, we are interested in studying the presence of stochastic stationary patterns.  We remember that stochastic Turing patterns \cite{Goldenfeld,Biancalani_2010} are signaled by the presence of at least a peak for the power spectrum in the direction of $k$, the spatial wavenumber, for $\omega=0$, where $\omega$ stands for the time frequency.  We are therefore going to analyze the functions $P_{Z}(k,0) \equiv P_1(k,0)$ and $P_{Y}(k,0)\equiv P_2(k,0)$, which respectively reads:

\begin{equation}
P_{Z}(k,0)=\dfrac{C_{Z,k}}{\Omega^{4}_{0}}=\frac{b_{22}(\mathcal{J}_{11}+\delta \tilde{\triangle})^{2}+(b_{11}-2\hat{\phi}\delta \tilde{\triangle}){\mathcal{J}_{21}}^2}{(\det{\mathcal{J}}+\mathcal{J}_{22}\delta  \tilde{\triangle})^2}\label{Pzfin}
\end{equation}
\begin{equation}
P_{Y}(k,0)=\dfrac{C_{Y,k}}{\Omega^{4}_{0}}=\frac{(b_{11}-2\hat{\phi}\delta \tilde{\triangle}){\mathcal{J}_{22}}^2+{b_{22}{\mathcal{J}_{12}}^2}}{(\det{\mathcal{J}}+\mathcal{J}_{22}\delta \tilde{\triangle})^2}, \label{Pyfin}
\end{equation}

where we have introduced:

\begin{eqnarray}
b_{11} &=& f_{1}(\hat{\phi},\hat{\psi})+f_{2}(\hat{\phi},\hat{\psi}), \\ 
b_{22}&=& g_{1}(\hat{\phi},\hat{\psi})+g_{2}(\hat{\phi},\hat{\psi}).
\label{par_b}
\end{eqnarray}

To study the conditions that yield to one or more peaks, we need to calculate the power spectrum derivative. We make use of the notation  $g(k) \equiv \delta \tilde{\triangle}=2\delta(\cos{k}-1)$ and obtain the following general expression:
\begin{equation}
\frac{dP_{j}(k,0)}{dk}=\frac{g'(k)}{(\det{\mathcal{J}}+\mathcal{J}_{22}g(k))^3}\{B_{j}g(k)+C_{j}\} 
\qquad \text{for} \quad j\in{\{Z,Y\}},
\label{derpriYZ}
\end{equation}
where $B_j$ and $C_j$ are defined as:
\begin{align}
B_Z&=2\hat{\phi}\mathcal{J}_{22}^3\\
C_Z&=-2\mathcal{J}_{22}\left(b_{11}\mathcal{J}_{22}^2+b_{22}\mathcal{J}_{12}^2+\hat{\phi}\mathcal{J}_{22}\det{\mathcal{J}}\right)\\
B_Y&=2\mathcal{J}_{21}(-b_{22}\mathcal{J}_{12}+\hat{\phi}\mathcal{J}_{21}\mathcal{J}_{22})\\
C_Y&=-2\mathcal{J}_{21}(b_{22}\mathcal{J}_{11}\mathcal{J}_{12}+\hat{\phi}\mathcal{J}_{21}\det{\mathcal{J}}+b_{11}\mathcal{J}_{22}\mathcal{J}_{21}).
\end{align}

Recall that $J_{ij}$ are the entries of the Jacobian matrix of system $(\phi_i,\psi_i) \mapsto (f_1-f_2,g_1-g_2)$ and $b_{ij}$ are given by eqs. (\ref{par_b}). 

We observe that  $k=0$ and $k=\pi$ are always stationary points of $P_j$. In fact $g'(k)=-2\delta \sin(k)$ is null if $k=0, \pi$. 
To have additional stationary points of $P_j$, one should require the quantity $B_{j}g(k)+C_{j}$ to vanish. 
This implies:
$$
 \cos(k)=1-\frac{C_{j}}{2\delta B_{j}}.
$$  
As $\cos(k) \in [-1,1]$,  it is necessarily the case that:
\begin{equation}
0\leqslant\frac{C_{j}}{2\delta B_{j}}\leqslant2.
\label{limitazione}
\end{equation}
Then, the derivative of $P_j$ can be zero in $k$ if $B_j$ and $C_j$ have the same sign. 
We indicate as $k_1$ and $k_2$, the stationary wavenumbers different from $\pi$.
\par
There are only two possible cases for the existence of $k_1$ and $k_2$:
\\[12pt]
\fbox{\begin{minipage}[b]{15.3 cm}\textbf{(i) Existence condition of} $\mathbf{k_1}$, $\mathbf{k_2}$
\begin{enumerate}[(a)]
      \item ${B_{j},C_{j}>0}\quad \text{and} \quad {\delta \geqslant \tfrac{C_{j}}{4B_{j}}}$, 
      \item ${B_{j},C_{j}<0}\quad \text{and} \quad {\delta \geqslant \tfrac{|C_{j}|}{4|B_{j}|}}$.
\end{enumerate}\end{minipage}}
\\[12pt]
We are interested to know whether $k_1$ and $k_2$ correspond to maxima or minima of $P_{j}(k,0)$. 
To achieve this goal we calculate the second derivative of $P_{j}(k,0)$:
\begin{equation}
{\frac{d^{2}}{dk^2}P_{j}(k,0)}=\frac{g''(k)\left(B_{j}g(k)+C_{j}\right)+B_{j}g'(k)^2}{(\det{\mathcal{J}}+\mathcal{J}_{22}g(k))^3}-\frac{3\mathcal{J}_{22}g'(k)^2\left(B_{j}g(k)+C_{j}\right)}{(\det{\mathcal{J}}+\mathcal{J}_{22}g(k))^4}.
\label{dersecPy}
\end{equation}
Remember that $k_1$ and $k_2$ are solution of $B_{j}g(k)+C_{j}=0$. 
The expression of the second order derivative is therefore cast into the form:
\begin{equation}
{\frac{d^{2}}{dk^2}P_{j}(k,0)}{\bigg{|}_{k=k_1,k_2}}=\frac{B_{j}g'(k)^2}{(\det{\mathcal{J}}+\mathcal{J}_{22}g(k))^3}.
\label{dersecPyk1}
\end{equation}
The nature of the stationary points $k_1$ and $k_2$ depends on the sign of both the denominator and $B_j$ in
\eqref{dersecPyk1}.
In particular, if we require that the points are maxima, or equivalently the second derivative in $ k_1 $ and $ k_2 $ 
has a negative sign, we must check one of the two following conditions: 
\\[12pt]
\fbox{\begin{minipage}[b]{15.3 cm}\textbf{(ii) Maximum conditions for points} $\mathbf{k_1}$, $\mathbf{k_2}$
\begin{enumerate}[(a)]
      \item ${B_{j}<0}\quad \text{and} \quad {\det{\mathcal{J}}+\mathcal{J}_{22}g(k)_{\big{|}_{k=k_1,k_2}}>0}$, 
      \item ${B_{j}>0}\quad \text{and} \quad {\det{\mathcal{J}}+\mathcal{J}_{22}g(k)_{\big{|}_{k=k_1,k_2}}<0}$.
\end{enumerate}\end{minipage}}
\\[12pt]
As anticipated we shall consider the case of a self-inhibitory non mobile species, which corresponds to requiring
$\mathcal{J}_{22}<0$. The denominator in \eqref{dersecPyk1} is then always positive, while $g(k)$ is by definition negative.
Accordingly, the kind of stationary points $k_1$ and $k_2$ depend on the sign of $B_j$. In particular, for the condition of maximum \textbf{(ii)}, ${{B_j}}$ must be negative. 
\par
To characterize whether the other stationary points $0, \pi$ are maxima or minima, we should again turn to evaluating the 
second derivatives for such choices of $k$.
As $g'(0)=0$, then equation \eqref{dersecPy} is:
\begin{equation}
\label{dersecPy0}
{\frac{d^{2}}{dk^2}P_{j}(k,0)}{\bigg{|}_{k=0}}=\frac{g''(0)\left(B_{j}g(0)+C_{j}\right)}{(\det{\mathcal{J}}+\mathcal{J}_{22}g(\pi))^3}
=\frac{-2\delta C_{j}}{(\det{\mathcal{J}}-4\delta \mathcal{J}_{22})^3}.
\end{equation}
Therefore $k=0$ is a maximum, if one of the following conditions is true:
\\[12pt]
\fbox{\begin{minipage}[b]{15.3 cm}\textbf{(iii) Maximum condition for} $\mathbf{k=0}$
\begin{equation*}
\text{(a)}\left\{
\begin{alignedat}{2}
& {-2\delta C_{j}<0}\\
& {(\det{\mathcal{J}}-4\delta \mathcal{J}_{22})>0}.
\end{alignedat}
\right.
\qquad \qquad
\text{(b)} \left\{
\begin{alignedat}{2}
& -2\delta C_{j}>0\\
& {(\det{\mathcal{J}}-4\delta \mathcal{J}_{22})<0}.
\end{alignedat}
\right.
\end{equation*}
\end{minipage}}
\\[12pt]
Since by assumption $\mathcal{J}_{22}<0$, condition \textbf{(iii)}(b) cannot be met. 
This is because the quantity $\det{\mathcal{J}}-4\delta \mathcal{J}_{22}$ is positive, as $\det{\mathcal{J}}>0$ since 
we have assumed that  $(\hat{\phi},\hat{\psi})$ is a stationary stable fixed point. 
The nature of the stationary point $k=0$ ultimately depends on the sign of $C_j$. 
If $C_j>0$, it is a maximum point, while, if $C_j<0$, it is a minimum.
\par
Consider now $k=\pi$ and observe that $g'(\pi)=0$. Equation \eqref{dersecPy} reads:
\begin{equation}
\label{dersecPypi}
{\frac{d^{2}}{dk^2}P_{j}(k,0)}{\bigg{|}_{k=\pi}}=\frac{g''(\pi)\left(B_{j}g(\pi)+C_{j}\right)}{(\det{\mathcal{J}}+\mathcal{J}_{22}g(\pi))^3}
=\frac{2\delta\left(-4\delta B_{j}+C_{j}\right)}{(\det{\mathcal{J}}-4\delta \mathcal{J}_{22})^3}
\end{equation}
For having a maximum in $k=\pi$ one of the following conditions must be satisfied:
\\[12pt]
\fbox{\begin{minipage}[b]{15.3 cm}\textbf{(iv) Maximum condition for} $\mathbf{k=\pi}$
\begin{equation*}
\text{(a)} \left\{
\begin{alignedat}{2}
& {-4\delta B_{j}+C_{j}<0}\\
& {(\det{\mathcal{J}}-4\delta \mathcal{J}_{22})>0}.
\end{alignedat}
\right.
\qquad \qquad
\text{(b)} \left\{
\begin{alignedat}{2}
& -4\delta B_{j}+C_{j}>0\\
& {(\det{\mathcal{J}}-4\delta \mathcal{J}_{22})<0}.
\end{alignedat}
\right.
\end{equation*}
\end{minipage}}
\\[12pt]
Since $\mathcal{J}_{22}<0$, the condition \textbf{(iii)}(b) is never satisfied: as already remarked, the term 
$\det{\mathcal{J}}-4\delta \mathcal{J}_{22}$ is in fact always positive. 
\par
\smallskip
Notice that, if $k=\pi$ is a maximum the values $k=k_1$ and $k=k_2$ are minima. 
Otherwise if $k_1$ and $k_2$ are maxima, $k=\pi$ is a minimum.
To show this, let us consider two different cases, respectively $B_{j}<0$ and $B_{j}>0$.
\par
\smallskip
If $B_{j}<0$ and, at the same time, condition \textbf{(i)} is satisfied, then $k_1$ e $k_2$ exist. 
In this case, the condition \textbf{(ii)}(a) guarantees that the stationary points else than $\pi$ are maxima. 
Indeed, $B_{j}<0$ and $(\det{\mathcal{J}}+\mathcal{J}_{22}g(k))_{\big{|}_{k=k_1,k_2}}$ is positive.
The condition for having a maximum in $k=\pi$, namely $-4\delta B_{j}+C_{j}<0$, is in contradiction with \textbf{(i)}. 
If $B_{j},C_{j}<0$, in fact, we can write $-4\delta B_{j}+C_{j}<0$. 
Taking into account the signs of the quantities involved, it results $ 4 \delta | B_ {j} | - | C_ {j} | <0 $, which implies 
$\delta <\frac {| C_ {j} |} {4 | B_ {j} |} $, in disagreement with the condition \textbf {(i)}. 
In conclusion $k=\pi$ is necessarily a minimum.
\par
\smallskip
Let us now turn to considering the case $B_{j}>0$. 
To have the existence of $k_1$ and $k_2$ one must impose $C_{j}>0$ and $\delta>\frac{C_{j}}{4B_{j}}$. 
Clearly, condition \textbf{(ii)} cannot be then satisfied and the two stationary points are minima. 
A maximum is instead found in $k=\pi$, as dictated by condition \textbf{(iv)}(b). 
\par
\medskip

A summary of the above results is given in the Tables annexed below, where the different scenarios are highlighted depending on the sign of the reference quantities. We recall that our results have been derived under the hypothesis of discrete lattice spacing $a$ (set to one in the calculations).  
Similar Tables can be in principle obtained for the case of a spatially continuum lattice, i.e. when $a \rightarrow 0$  and $g(k) \equiv - \delta k^2$. It can be however shown  \cite{McKane,De_Anna} that the power spectrum of fluctuations scales with an amplitude prefactor proportional to $a^d$, $d$ being the dimension of the embedding space ($d=1$, in our case). Hence, in the limit $a \rightarrow 0$, fluctuations fade away and the stochastic pattering is non detectable. However, as remarked in \cite{Goldenfeld},
another continuum limit can be performed, starting from the same microscopic discrete formulation. 
One could in fact imagine to keep patch dimension to a constant, while sending to infinity both $\omega$ and the linear size of the physical space which hosts the system under scrutiny. This is indeed the case considered in \cite{Biancalani_waves}: working under this alternative scenario, fluctuations, and so the triggered patterns, are persistent also in the continuum limit. The choice of operating with patches of finite size, where microscopic constituents are supposed well mixed, and accounting for the possibility of jumping towards neighbor patches of a finite lattice, proves useful when modeling ecological systems  \cite{McKaneEcologyTrends},  
or in cellular biology, the space inside the membrane being partitioned in macro compartments and oganelles \cite{cell},  but also for studying chemical systems as e.g. the device introduced in \cite{solomon}.

\begin{tabular}{ |r|c|c| }
\hline
\multicolumn{1}{|r|}{$\mathbf{\mathcal{J}_{22}<0}$}
 &  \multicolumn{1}{|c|}{ $\mathbf{C_{j}>0} $}\\
\hline
$\mathbf{B_j>0}$  &
  $\delta\geqslant\frac{C_{j}}{4B_{j}}\quad\exists\quad k_1$ and $k_2$ and are minima. Maxima are found in $k=0,\pi,2\pi$\\
 
 & \includegraphics[scale=0.4]{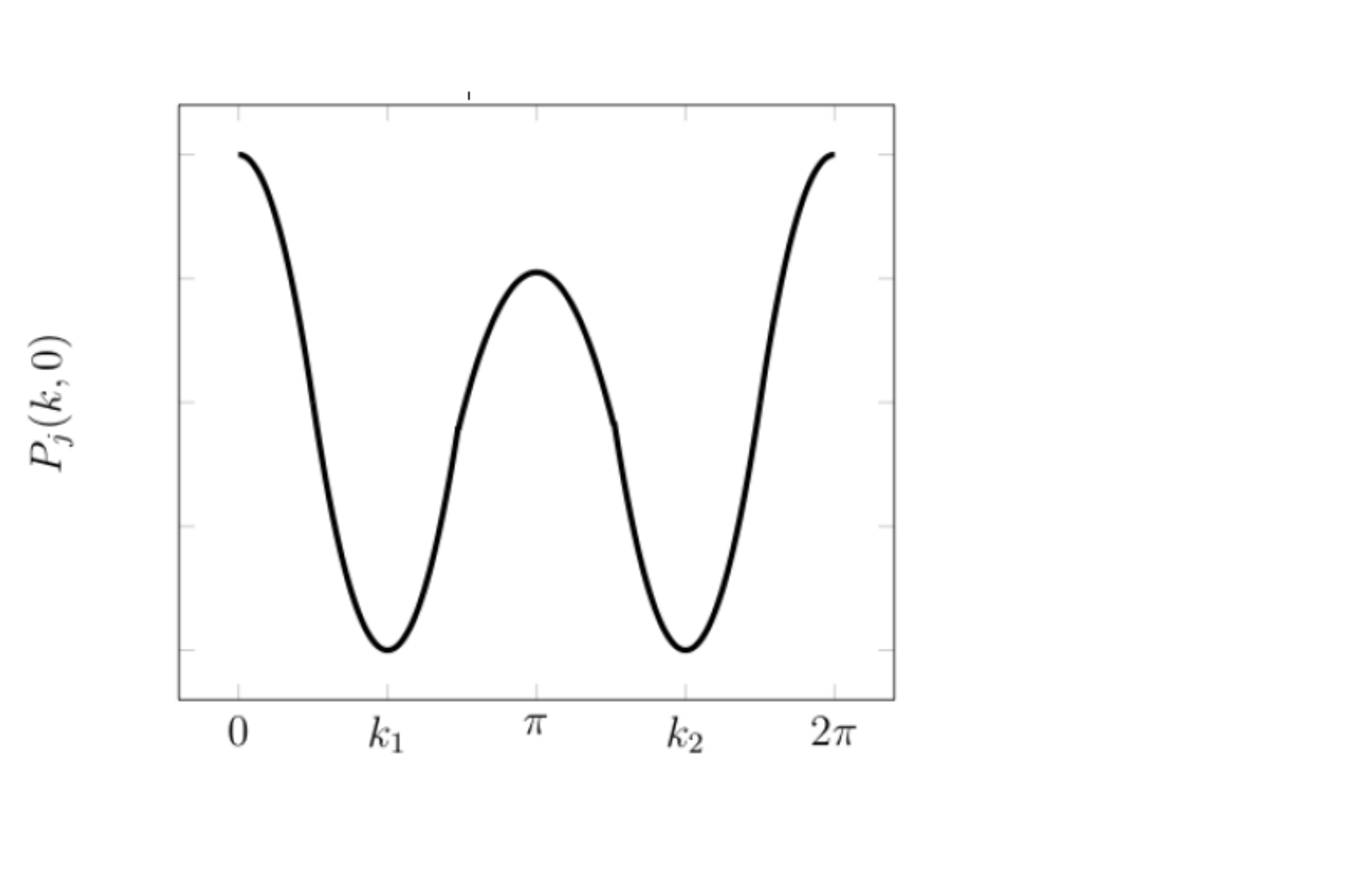} \\

 & $\delta<\frac{C_{j}}{4B_{j}}\quad\not\exists\quad k_1$ and $k_2$. $k=0$ and $k=2\pi$ are maxima. A minimum is found in $k=\pi$.\\
 
 & \includegraphics[scale=0.4]{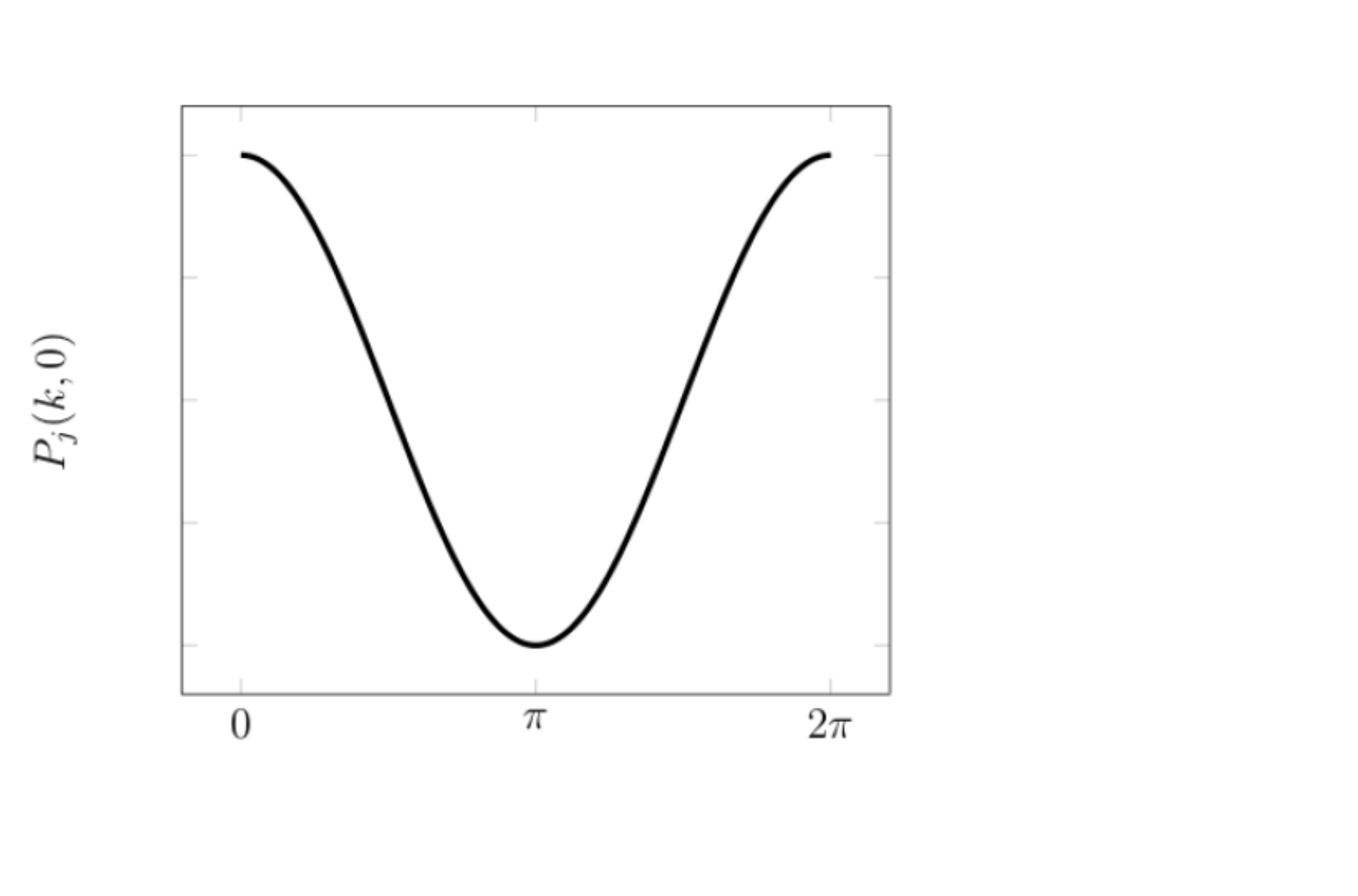} \\
\hline

$\mathbf{B_j<0}$ & $\not\exists\quad k_1$ and $k_2$. $k=0$ and $k=2\pi$ are maxima. A minimum is found in $k=\pi$.\\

& \includegraphics[scale=0.4]{fig_T4.pdf}\\ 
\hline
\end{tabular}

\begin{tabular}{ |r|c|c| }
\hline
\multicolumn{1}{|r|}{$\mathbf{\mathcal{J}_{22}<0}$}
 &  \multicolumn{1}{|c|}{ $\mathbf{C_{j}<0} $}\\
\hline
$\mathbf{B_j>0}$  & $\not\exists\quad k_1$ and $k_2$. $k=\pi$ is always a maximum. Two minima are found in $k=0$ and $k=2\pi$\\
 
 & \includegraphics[scale=0.4]{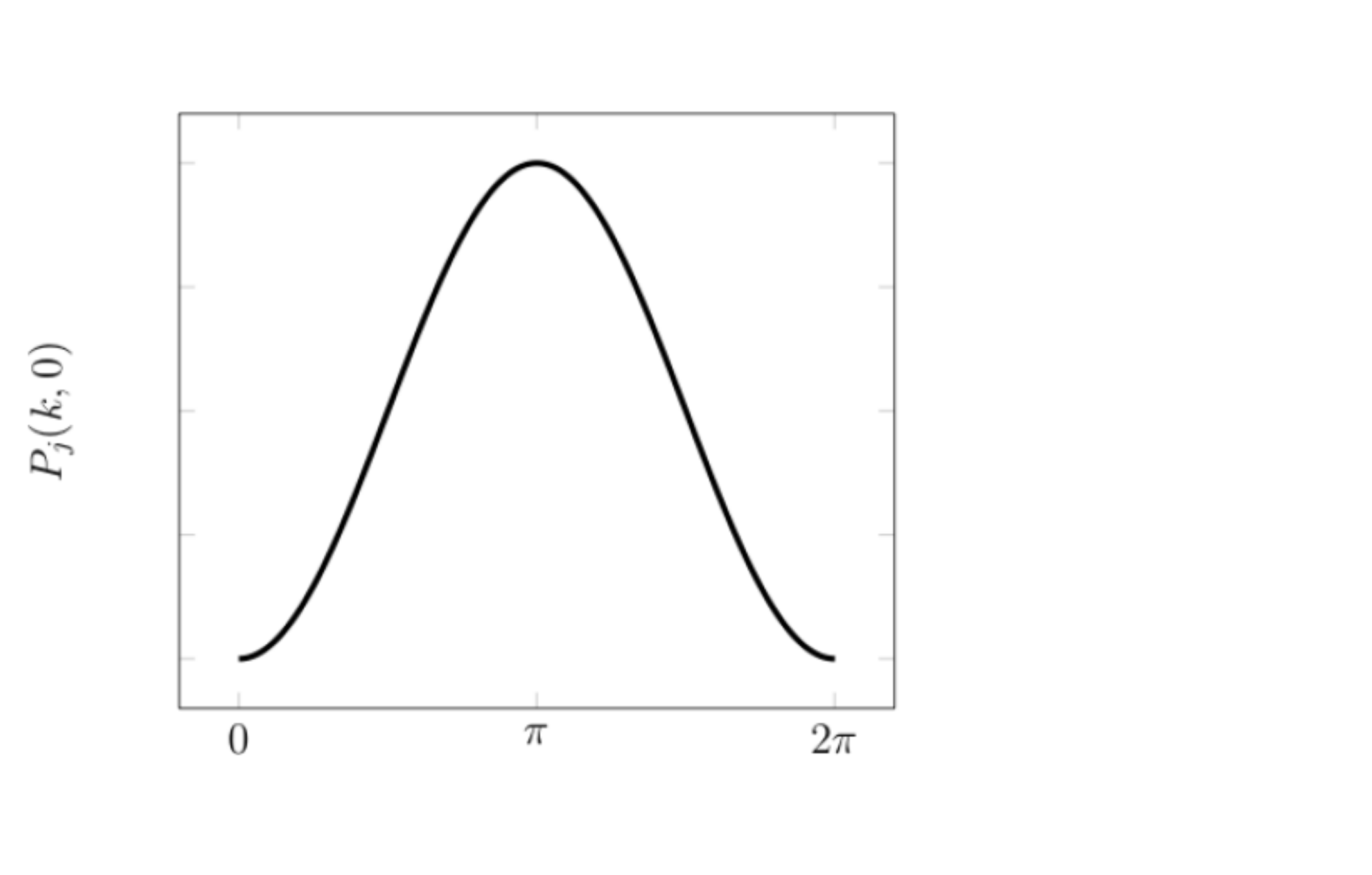} \\

\hline
$\mathbf{B_j<0}$ & $\delta>\frac{C_{j}}{4B_{j}}\quad\exists\quad k_1$ and $k_2$ and are maxima. $k=0,\pi,2\pi$ are minima.\\
 
 & \includegraphics[scale=0.4]{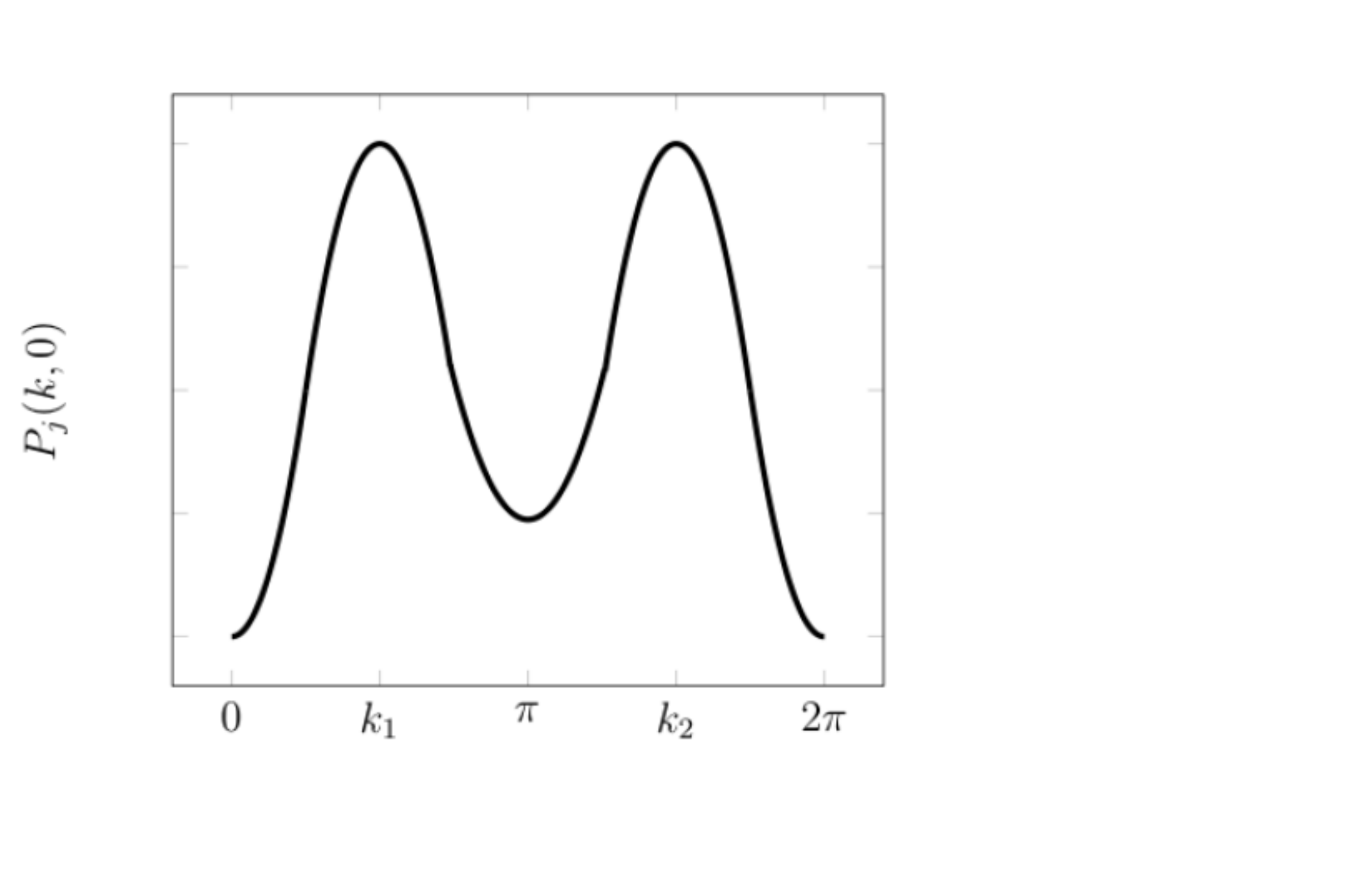} \\

& $\delta<\frac{C_{j}}{4B_{j}}\quad\not\exists\quad k_1$ and $k_2$. $k=0$ and $k=2\pi$ are minima. A maximum is found in $k=\pi$.\\

& \includegraphics[scale=0.4]{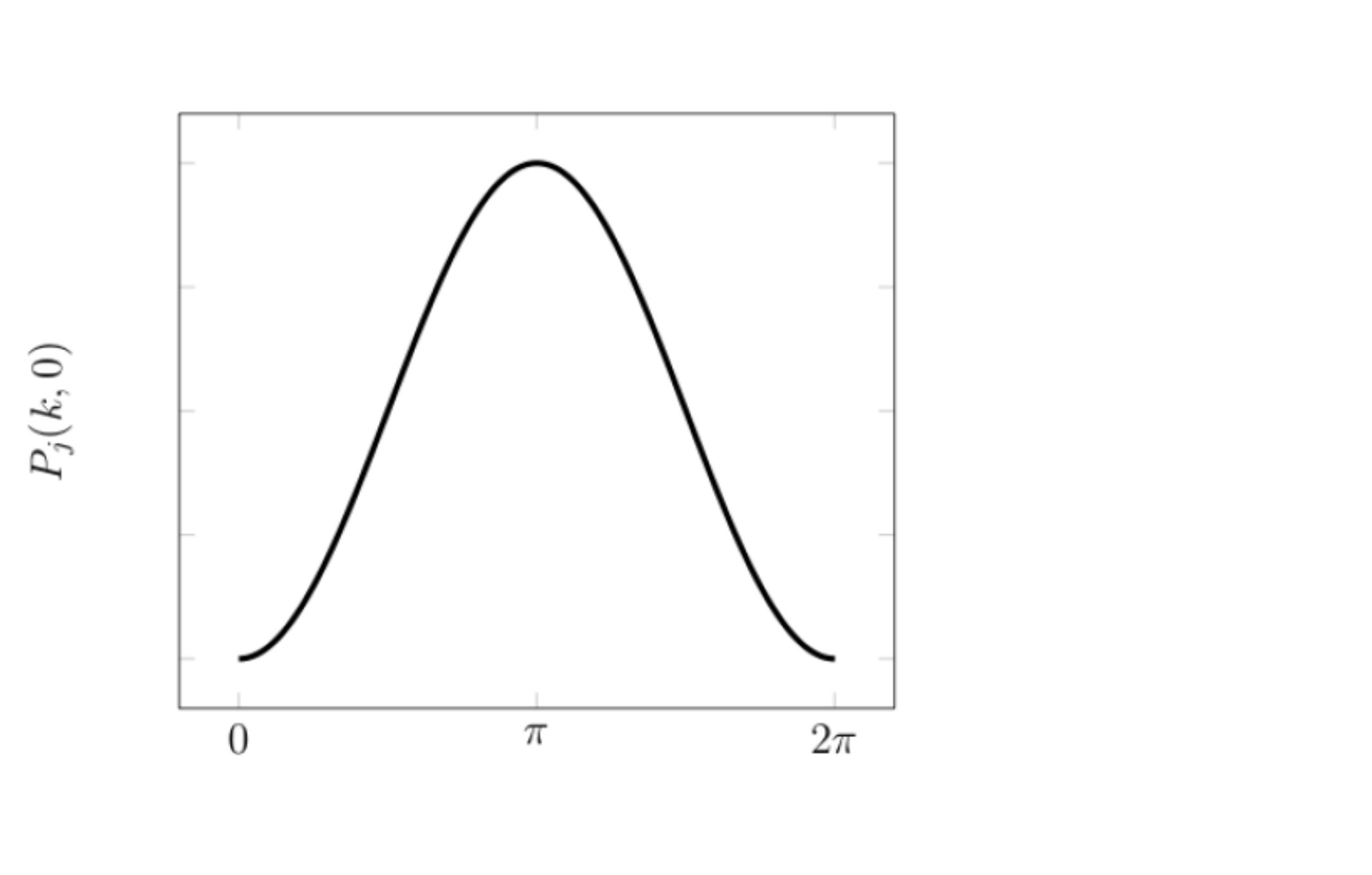}\\ 
\hline
\end{tabular}


\newpage
\section{A simple stochastic reaction--diffusion model}  \label{S6}

We have so far demonstrated that stochastic Turing patterns exist for reaction-diffusion models, defined on a discrete lattice, in which only one species diffuses. 
Working in a general context, we elaborated on the conditions which lead to Turing pattern, mediated by demographic noise. 
\par
As an application of the results discussed above, we consider a specific stochastic reaction-diffusion model, which can be cast 
in the form specified by (\ref{EC1}) and (\ref{EC2}). 
We choose in particular:
\begin{eqnarray}
f_1\bigg(\frac{s_i}{V},\frac{q_i}{V}\bigg)&=&\eta_1 \\
f_2\bigg(\frac{s_i}{V},\frac{q_i}{V}\bigg)&=&\eta_2 \left(\frac{s_i}{V}\right)^{p}+
\eta_3 \left( \frac{q_i}{V} \right)^{n} \\
g_1\bigg(\frac{s_i}{V},\frac{q_i}{V}\bigg) &=& \eta_4 \\
g_2\bigg(\frac{s_i}{V},\frac{q_i}{V}\bigg) &=& \eta_5 \left(\frac{s_i}{V}\right)^{p}+
\eta_6 \left( \frac{q_i}{V} \right)^{n}
\end{eqnarray}
to define the microscopic reaction rates implicated in chemical equations (\ref{EC1}). 
Here $\eta_i$ are positive real numbers, while $p$ and $t$ are integers. We will set $p=4$ and $n=1$. 
Note that the proposed model has no specific applied interest: it is solely introduced for demonstrative purposes, 
aiming at testing the validity of the mathematical analysis developed above.  

In the mean-field approximation, one gets: 
\begin{equation}
\left\{
\begin{alignedat}{1}
&\frac{\partial \phi_{i}}{\partial t}=-\eta_2 \phi_i^{p}
-\eta_3 \psi_i^{n} + \eta_1 + \delta   \Delta \phi_i \\
&\frac{\partial \psi_{i}}{\partial t}=-\eta_5 \phi_i^{p}
-\eta_6 \psi_i^{n} +\eta_4\,.
\label{R-D1d_appl}
\end{alignedat}
\right.
\end{equation}
To calculate homogeneous fixed point $(\hat{\phi}, \hat{\psi})$  of system (\ref{R-D1d_appl}) one needs 
needs to solve the following equations: 

\begin{equation}
\left\{
\begin{alignedat}{1}
-\eta_2 \hat{\phi}^{p}
-\eta_3 \hat{\psi}^{n} + \eta_1 =0\\
-\eta_5 \hat{\phi}^{p} -\eta_6 \hat{\psi}^{n} +\eta_4 = 0 \,.
\label{R-D1d_appl_1}
\end{alignedat}
\right.
\end{equation}
which immediately yield:

\begin{eqnarray}
\label{fix_p_1}
\hat{\phi} &=&  \left(\frac{\eta_1 \eta_6-\eta_3 \eta_4}{\eta_2 \eta_6-\eta_3 \eta_5}\right)^{1/p} \\
\hat{\psi} &=& \left(\frac{\eta_2 \eta_4-\eta_1 \eta_5}{\eta_2 \eta_6-\eta_3 \eta_5} \right)^{1/n} 
\end{eqnarray}

The parameters are to be in turn assigned so that the above fixed point is real and positive, a condition on which we shall return in the following.  
Furthermore, we require $(\hat{\phi}, \hat{\psi})$ to be a stable fixed point, so to match the theory prescriptions. The trace of the Jacobian matrix $\mathcal{J}$ associated to the homogeneous (a-spatial) version of system (\ref{R-D1d_appl}) reads:

\begin{equation}
\Tr (\mathcal{J}) = - \left( \eta_2 p \hat{\phi}^{p-1} + \eta_6 n \hat{\psi}^{n-1} \right).
\end{equation}

The trace is therefore always negative, for any choice of the parameters which returns a physically sound ($\hat{\phi}, \hat{\psi}>0$) homogeneous fixed point.
For the fixed point to be stable, one should further impose:

\begin{equation}
\det(\mathcal{J}) = \left( \eta_2 \eta_6-\eta_3 \eta_5 \right) p n  \hat{\phi}^{p-1} \hat{\psi}^{n-1} > 0. 
\end{equation}

This latter condition translates in: 

\begin{equation}
\label{positive_det}
\eta_3 < \left( \frac{\eta_2}{\eta_5} \right) \eta_6 \equiv \gamma_1 \eta_6, 
\end{equation}

where we brought into evidence the dependence on $\eta_6$ and $\eta_3$, since they will later on act as control parameters. By using the above condition (\ref{positive_det}) 
into equations (\ref{fix_p_1}) the condition for positive concentrations $\hat{\phi}, \hat{\psi}>0$ gives:

\begin{eqnarray}
\label{positive_fix_point1}
&\eta_2& \eta_4-\eta_1 \eta_5 \equiv \gamma_2 >0 \\
&\eta_3& < \left( \frac{\eta_1}{\eta_4} \right) \eta_6 \equiv \gamma_3 \eta_6.
\end{eqnarray}   

The homogeneous fixed point ($\hat{\phi}, \hat{\psi}$) determined above exists and it is stable, provided conditions (\ref{positive_det}) and (\ref{positive_fix_point1}) are simultaneously met. Moreover, and 
as discussed in the first part of the paper, the spatially extended system (\ref{R-D1d_appl}) cannot experience a (deterministic) Turing instability since $g_{\psi} = -n \eta_6 \hat{\psi}^{n-1}$ is by definition negative. The homogeneous fixed point is hence a stable, although trivial attractor of the spatial deterministic model.   
\par
A different scenario holds instead when the stochastic version of the deterministic model (\ref{R-D1d_appl}) is considered.
As we will show, it is in fact possible to assign the model parameters so as to generate a power spectrum of the stochastic
fluctuations with two maxima for non trivial values of $k_1$ and $k_2$, for $\omega=0$. 
These maxima are interpreted as the signature of stochastic Turing patterns. 

\begin{figure}
\includegraphics[width=0.6\textwidth]{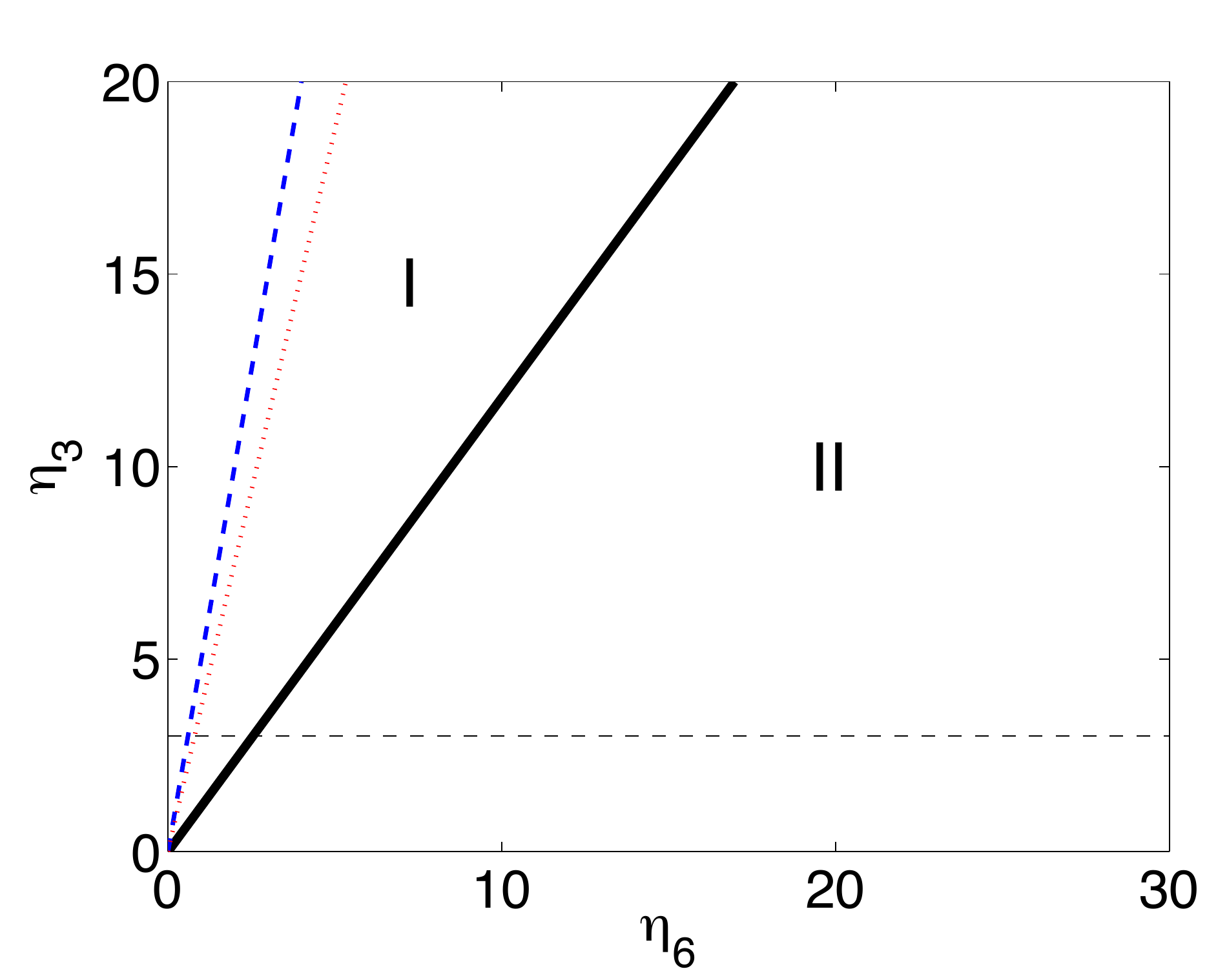}\\
  \caption{
  The plane ($\eta_6$, $\eta_3$) is partitioned into two regions. In region II, the power spectrum of fluctuations is predicted to display two peaks in, respectively, $k_1$ and $k_2$. These are positions symmetric with respect to $\pi$. In region I the power spectrum has instead a maximum in $k=0$. The parameters are $\eta_1=15$; $\eta_2=20$; $\eta_4=4$; $\eta_5=4$; $\delta=42.9473$. With this choice, $\gamma_1=4$; $\gamma_3=20>0$; $\gamma_3=3.75$. The two lines which cross the origin represent respectively the two conditions $\eta_3 = \gamma_1 \eta_6$ (blue online) and $\eta_3 = \gamma_3 \eta_6$ (red online). Region I is delimited by this latter and the thick solid line which marks the transition to the adjacent region II. The horizontal dashed lines is drawn at $\eta_3=3$: the data reported in the following figures (\ref{fig:simulations}) and (\ref{fig:bif}) refer to choices of the parameters that fall on such a line. 
}\label{zone}
\end{figure}

To this end we fix all parameters to nominal, arbitrarily chosen values, except for $\eta_3$ and $\eta_6$ which can be tuned. 
We will then adjust $\eta_3$ and $\eta_6$ so to match conditions {\bf (i)} and {\bf (ii)}, as outlined in the preceding section. 
This results in region II of the parameter plane, as depicted in Figure \ref{zone}. 
Conversely, in region I the power spectrum of fluctuations is predicted to display an isolated maximum for $k=0$.  

\begin{figure}
\begin{tabular}{cc}
\includegraphics[scale=0.33]{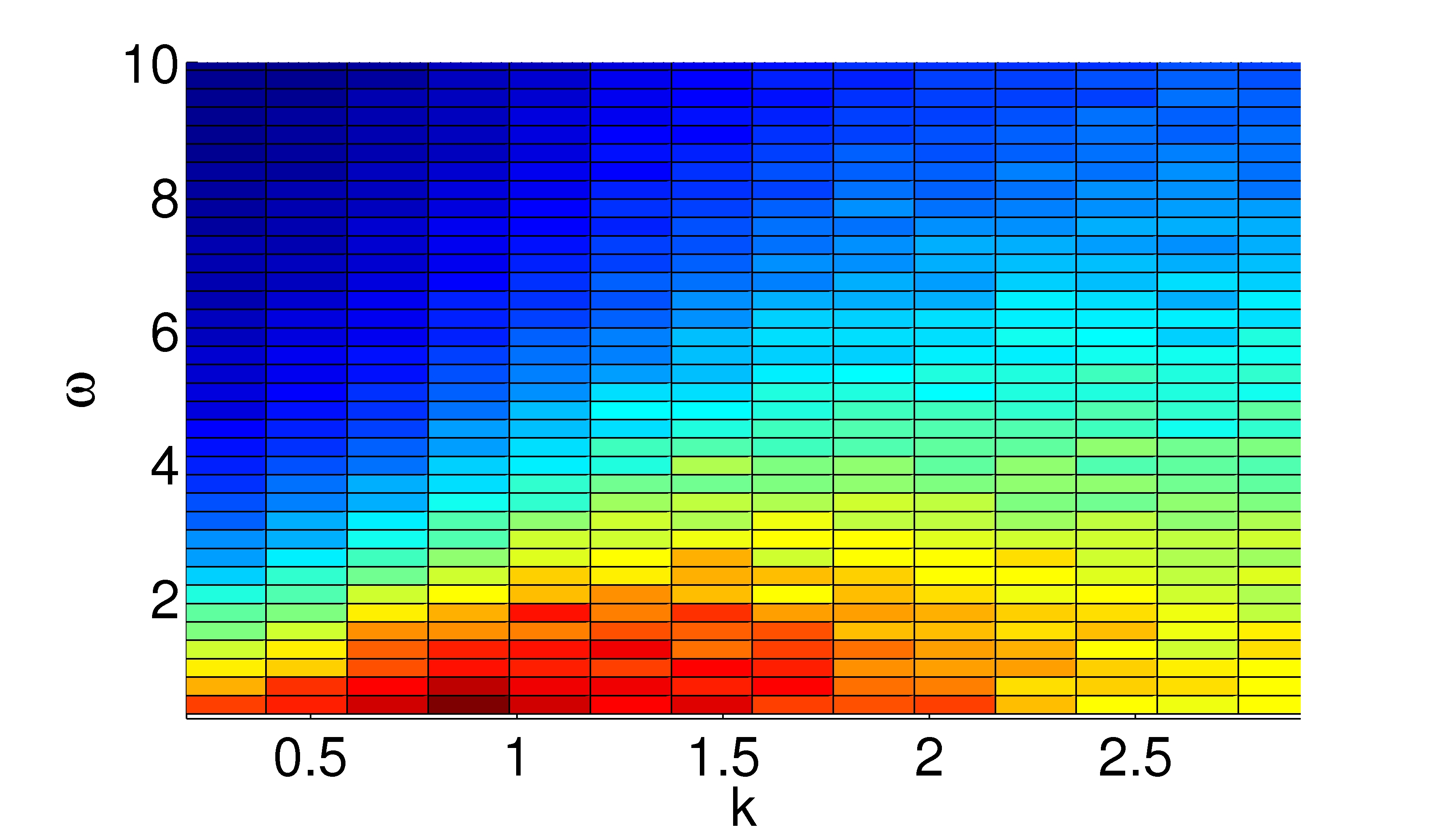}&
\includegraphics[scale=0.33]{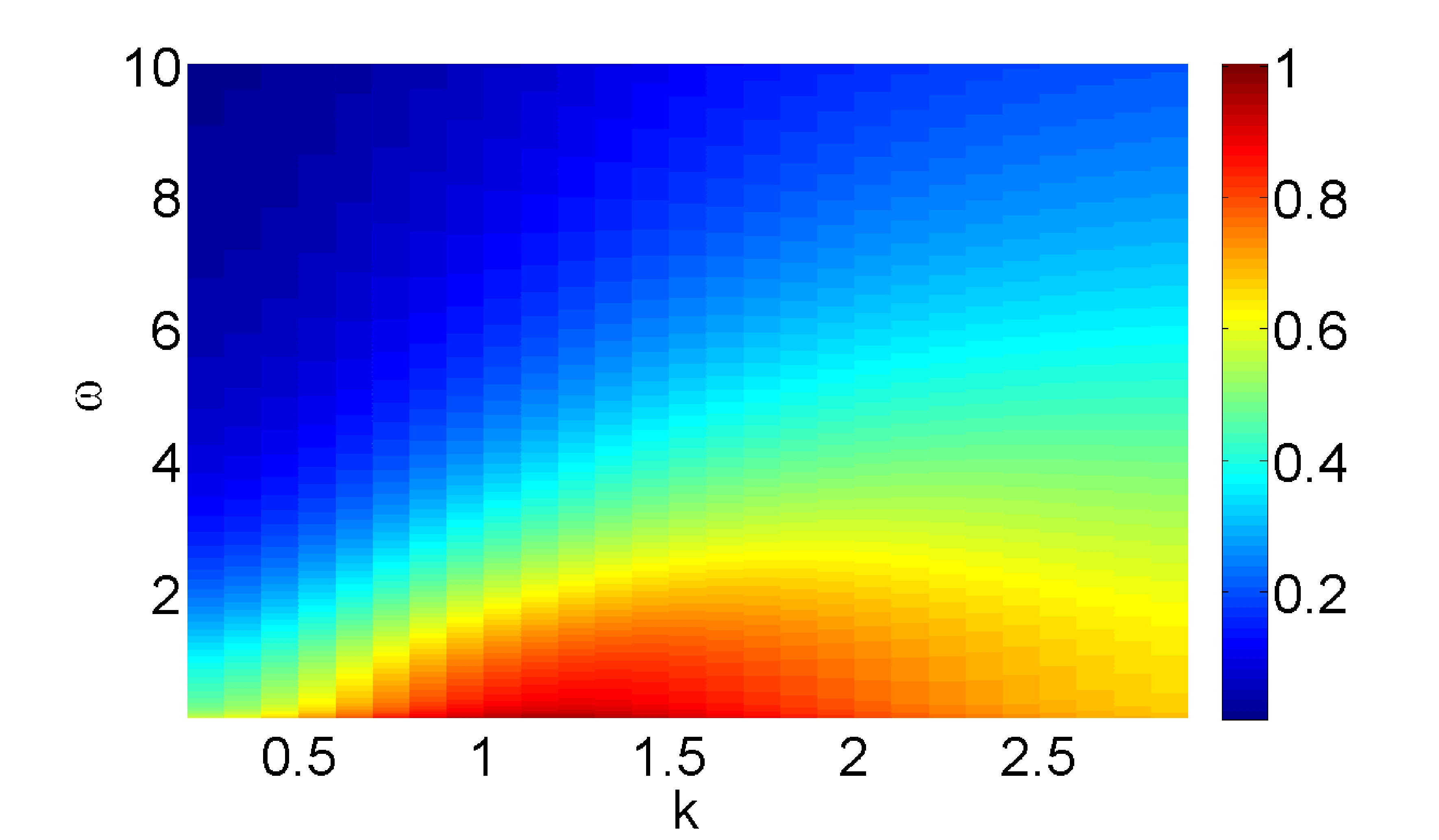}\\
(a) & (b)\\
\end{tabular}
\caption{In panel (a), the numerical power spectrum of the fluctuations for species $Z$ 
is represented, with an appropriate color code, in the plane ($\omega$, $k$), for a choice of the parameters that fall in region I of Figure \ref{zone}. Specifically, we have set
$\eta_6=25$, $\eta_3=3$. The other parameters are set to the values specified in the caption of Figure \ref{zone}. Here $V=5000$  and $\Omega=32$. The numerical power spectrum is obtained by averaging over $200$ independent realizations based on the Gillespie algorithm. A peak is found in the interval $[0,\pi]$. A symmetric maximum exists in $[\pi,2 \pi]$ (non displayed). In panel (b) the power spectrum calculated analytically is plotted and shown to agree with the numerical result.  The power spectra are normalized so to have maximum equal to unit. The color bar applies to both panels.  \label{fig:simulations}}
\end{figure}

In Figure \ref{fig:simulations}(b) we plot a two dimensional view of the theoretical power spectrum for a choice of the parameters
 ($\eta_6$, $\eta_3$) which falls in region II.  
The predicted profile is just displayed in the interval $k \in [0, \pi]$: a peak is present for a value of $k$ smaller than $\pi$. 
A second, specular, peak is clearly found for $k>\pi$. 
The two maxima of the power spectrum occur for $\omega=0$. 
They correspond therefore to stationary non homogeneous patterns. 
To validate the theory predictions we performed direct numerical simulations, by means of the Gillespie algorithm \cite{Gillespie}.
This is a Monte Carlo based scheme  which produces realizations of the stochastic dynamics equivalent to  those obtained from
 the governing master equation. 
The power spectrum calculated by averaging over a large collection of independent realizations of the stochastic dynamics is depicted in Figure \ref{fig:simulations}(a), showing a good agreement with the corresponding theoretical profile. 
This confirms the validity of the analysis developed above, and summarized in the Tables presented above. 

\begin{figure}
\includegraphics[scale=0.35]{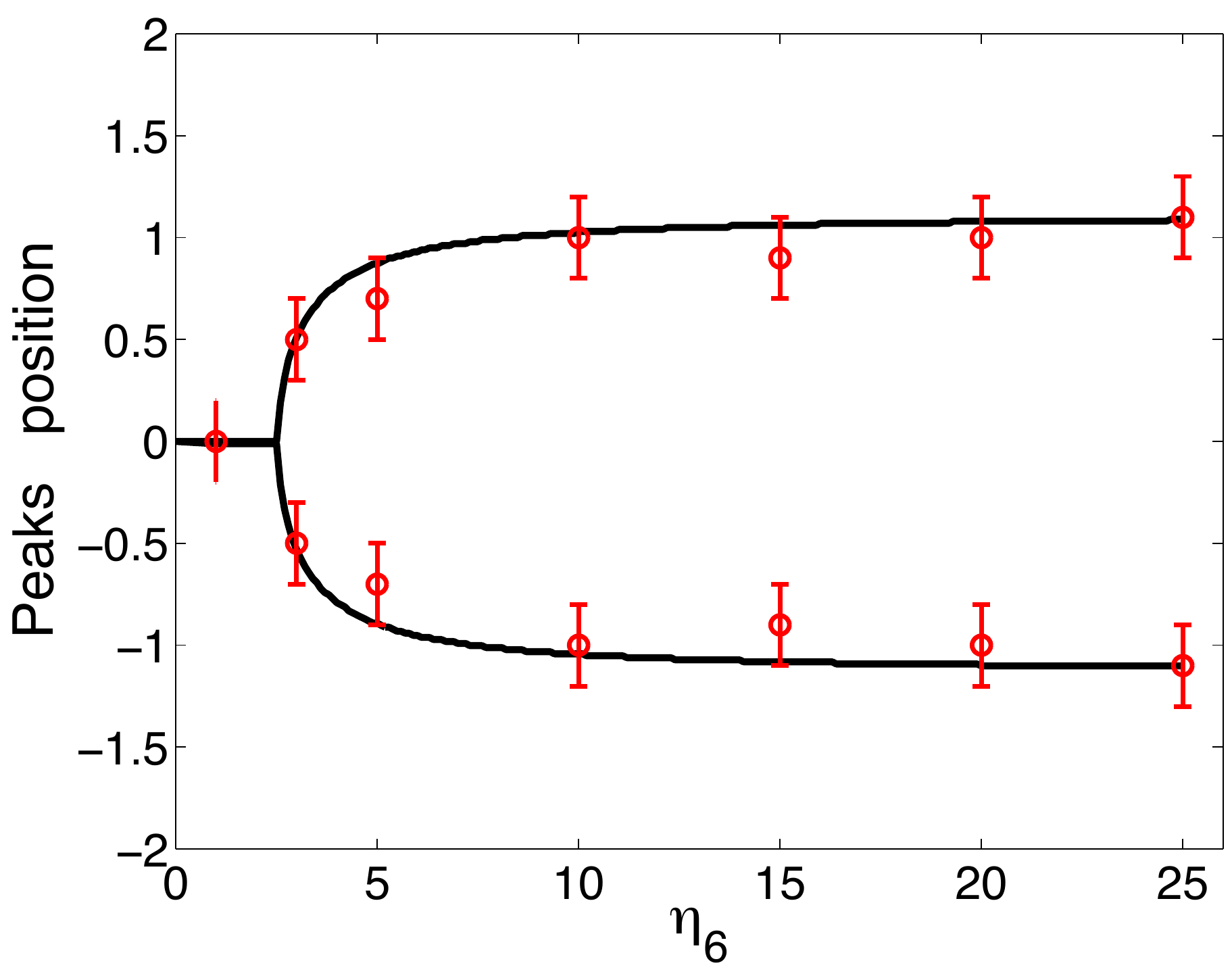}
\caption{A bifurcation diagram is displayed, which exemplifies the transition  from zone I to zone II. More specifically, the position of the peaks of the power 
spectrum of species $Z$ is plotted as a function of the control parameter $\eta_6$. Here, $\eta_3=3$, a value that corresponds to the horizontal dashed line in figure \ref{fig:simulations}. The solid line stands for the theory prediction, while the symbols refer to direct simulations of the stochastic dynamics. The simulations are averaged over $150$ independent realizations. The error in the location of the peak is assumed as twice the spacing of the imposed wavelength mesh. 
\label{fig:bif}}
\end{figure}

In figure \ref{fig:bif}, the position of the maxima of the power spectrum of species $Z$ is plotted as a function of the control parameter  $\eta_6$, while $\eta_3$ is set to the value that corresponds to the dashed horizontal line in figure \ref{zone}. This results in a  bifurcation diagram from zone I to zone II.  A similar plot can be obtained for the co-evolving species $Y$.  The solid line stands for the theoretical predictions, which follows the results summarized in the Tables annexed above. A transition from zone I (one isolated peak) to zone II (two symmetric peaks) is predicted to occur at $\eta_6 \simeq 2.5$. The symbols in figure \ref{zone} refer to the position of the power spectrum as obtained via direct simulations and confirms the correctness of the theoretical scenario.

A final comment is mandatory at this point. Fluctuations driven patterns are stochastic in nature: as such they are not stationary, unlike their deterministic analogue.  Stochastic patterns continuously decay, while they are recreated by the effect of the noise \cite{McKaneBiancalaniRoger}.  In general, the noisy nature of the patterns makes them hard to detect by visual inspection. The emergence of a length scale become often clear only via a Fourier analysis. This is the case for the simple model here 
investigated for demonstrative reasons: the patterns emerging from one single realization are indeed masked by a large amount of noise (data not shown). Similar conclusion are reached in \cite{woolley}  where stochastic simulations for the Schnakenberg kinetics \cite{schnak} are carried out just outside the (deterministic) region of Turing order. On the other hand, patterns can possibly become more distinct depending on the simulated model, the dimensionality of the system ($1D$ vs. $2D$ ) and the structure (lattice vs. network) of the embedding space. For the Levin-Segel model \cite{levin} studied in two dimension \cite{Goldenfeld}, stochastic patterns are quite visible at the naked eye. Similarly, robust and rather distinct patterns are found when a stochastic reaction model of the Brusselator type \cite{bruss} is defined on a network topology \cite{asslani}. Also, quasi--waves patterns found in \cite{Biancalani_waves} for a modified version of the Brusselator model with long range couplings, stand out rather clearly from one single realization of the stochastic dynamics.  The search for the necessary ingredients that make stochastic pattern accessible at visual inspection, remains however an important and still open question that deserves to be further addressed.

\section{Conclusion}  \label{S7}

Pattern formation is an important domain of study which finds many applications in distinct contexts of interest, including ecology, biology and chemistry. 
The Turing instability is one of the mechanisms that can be invoked to explain the emergence of stationary stable, spatially ordered patterns in reaction-diffusion models. 
These latter are systems of coupled partial differential equations which govern the time and space evolution of the continuum concentrations of constituents. 
As such, reaction diffusion models are deterministic in nature. 
They omit the stochastic contributions that need to be included when dealing with finite populations and, in this respect, represent an idealized approach to the modeling of the inspected 
phenomena. The classical, deterministic theory for the Turing instability requires that at least two species diffuse in a domain in which they are confined: the diffusion potentially leads to an instability in following a perturbation of a stable equilibrium of the homogeneous system. Conversely, if just one species is allowed to diffuse the Turing instability is always precluded, when the system is defined on a continuum support. 
Working on a discrete lattice, Turing patterns in principle develop, but just for a trivial choice of the most unstable wave number and limited to models that assume the non diffusing species to operate as a self-activator. 

Beyond the deterministic viewpoint, in the last few years the concept of stochastic Turing instability has been introduced in the literature \cite{Goldenfeld, Biancalani_2010}: discrete systems, made of a large though finite number of constitutive entities, can generate stochastic order on a macroscopic scale, as follows a resonant mechanism which self-consistently amplifies the intrinsic demographic noise. Elaborating on this concept, we 
considered a general stochastic reaction diffusion model, with just one diffusing species, and showed that stochastic Turing patterns are indeed possible also when the non mobile species has a self-inhibitory capability, i.e. a condition for which deterministic patterns are a priori excluded.  General analytical conditions for the existence of the stochastically driven patterns are given. The predictions are tested numerically working with a simplified model that falls in the general class of systems for which the theory has been developed.  The quantitative agreement observed between theory and simulations points to the validity of our analysis, which, we believe, could open up novel perspectives to tackle the problem of pattern formation beyond the classical deterministic picture.

\end{document}